\documentclass[fleqn,10pt]{wlscirep}
\usepackage[utf8]{inputenc}
\usepackage[T1]{fontenc}
\usepackage{csquotes}
\title{Static network structure cannot stabilize cooperation among Large Language Model agents}

\author[1]{Jin Han}
\author[2]{Balaraju Battu}
\author[3]{Ivan Romi\'{c}}
\author[2]{Talal Rahwan}
\author[1,3,*]{Petter Holme}
\affil[1]{Department of Computer Science, Aalto University, 00076 Aalto, Finland}
\affil[2]{New York University Abu Dhabi, United Arab Emirates}
\affil[3]{Center for Computational Social Science, Kobe University, Kobe, Hyogo 657-8501, Japan}

\affil[*]{petter.holme@aalto.fi}


\begin{abstract}
Large language models (LLMs) are increasingly used to model human social behavior, with recent research exploring their ability to simulate social dynamics. Here, we test whether LLMs mirror human behavior in social dilemmas, where individual and collective interests conflict. Humans generally cooperate more than expected in laboratory settings, showing less cooperation in well-mixed populations but more in fixed networks. In contrast, LLMs tend to exhibit greater cooperation in well-mixed settings. This raises a key question: Are LLMs about to emulate human behavior in cooperative dilemmas on networks? In this study, we examine networked interactions where agents repeatedly engage in the Prisoner’s Dilemma within both well-mixed and structured network configurations, aiming to identify parallels in cooperative behavior between LLMs and humans. Our findings indicate critical distinctions: while humans tend to cooperate more within structured networks, LLMs display increased cooperation mainly in well-mixed environments, with limited adjustment to networked contexts. Notably, LLM cooperation also varies across model types, illustrating the complexities of replicating human-like social adaptability in artificial agents. These results highlight a crucial gap: LLMs struggle to emulate the nuanced, adaptive social strategies humans deploy in fixed networks. Unlike human participants, LLMs do not alter their cooperative behavior in response to network structures or evolving social contexts, missing the reciprocity norms that humans adaptively employ. This limitation points to a fundamental need in future LLM design—to integrate a deeper comprehension of social norms, enabling more authentic modeling of human-like cooperation and adaptability in networked environments.
\end{abstract}

\begin{document}

\flushbottom
\maketitle

\thispagestyle{empty}

\section*{Introduction}

Over the last few years, the advances in artificial intelligence have ignited hopes of new methods for the behavioral and social sciences\cite{christakis_et_al,evans}. In particular, chatbots powered by so-called large language models (LLM)\cite{biever} are believed to be able to emulate humans in conversations well enough to eventually replace them in experiments\cite{argyle2023out}. Even though experiments with human participants are ultimately needed to advance the behavioral and social sciences, there are expectations AI agents could aid in experimental design and provide experimental agents with programmable behavior (so-called ``digital twins'') \cite{christakis_et_al,evans}. LLMs are trained on extensive datasets of human-generated text and semantic knowledge from various societies  \cite{christakis_et_al,argyle2023out}. These models operate as conditional probability distributions, where altering the context or narrative can steer them toward more desirable outcomes by influencing the likelihood of specific responses while reducing others \cite{argyle2023out}. Consequently, LLMs are highly skilled at following instructions and embodying assigned personalities \cite{mei2024turing}. When given personalities, LLMs can display traits that resemble human nature, almost as if they possess their mind  \cite{peters2024large,wang2024large}. This training process can also improve LLMs' understanding and reasoning regarding cooperation, defection, and balancing individual and collective interests \cite{lore2023strategic}.

Recent behavioral experiments have shown that LLM agents can effectively substitute human participants in certain contexts, particularly within Western, Educated, Industrialized, Rich, and Democratic (WEIRD) societies  \cite{argyle2023out}. However, this substitution does not extend to other cultural contexts  \cite{atari2023humans}. Beyond cultural differences, within the same cultures, individuals' social and strategic preferences depend on the network they are part of  \cite{shi2020freedom, centola2015spontaneous,airoldi2024induction,rand2014static}. Individuals can infer underlying network structures and adapt their social behavior, even without a complete overview of the network. Through interactions with neighbors and social learning, individuals discern the network they are part of, and this structure significantly influences their social behavior. 
A fascinating question arises: can large language models discern these network structures and adjust their behavior similarly to humans? This inquiry becomes especially intriguing when we consider their potential to help solve societal challenges, like promoting cooperation in social dilemmas. For this vision to come to life, it is crucial that AI agents consistently demonstrate behavior akin to that of humans.

We have some understanding of LLMs' capabilities in repeated prisoners' dilemma games in well-mixed populations  \cite{mei2024turing,akata_etal,lore2023strategic,lore_heydari,brookins_debacker,fontana2024nicer}, but we lack insight into how they perform in network settings. There, individuals consider not only strategies but also their interactions with specific network members  \cite{SZABO200797,santos2006cooperation}. Previous research has shown that individuals can establish cooperation when they have the opportunity to adjust their social ties based on their experiences with neighbors  \cite{santos2006cooperation}. Interestingly, it has been found that in a prisoner's dilemma game, individuals can achieve stable cooperation within specific static network structures when the benefit-to-cost ratio exceeds the number of connections. However, the same parameter settings do not foster cooperation in a well-mixed population. The findings suggest that when individuals are aware of their neighborhood and the consequences of their cooperative actions, they can adapt their strategies based on the success of their neighbors, leading to stable cooperation even in networks with a few defectors  \cite{rand2014static}.

In this study, we investigate whether LLMs can reliably adjust their social behavior in response to the network structure they operate within, as humans do. We study the capabilities of LLM agents, as of 2024, in the context of networked social dilemmas \cite{SZABO200797, goyal, mei2024turing}. These stylized situations are designed to explore how humans prioritize between short-sighted egoistic and long-term prosocial choices. Such dilemmas have a wide range of social science applications, from behavioral economics, which explores questions like how to best incentivize people to follow policies, to anthropology and evolution, which search for the unique behavioral mechanisms behind human cooperation among known lifeforms.


\section*{Methods}

\subsection*{The prisoner's dilemma}

The prisoner's dilemma is a game-theoretic model of a social dilemma \cite{camerer1997progress}. Cooperation involves offering a benefit ($b$) to another individual while incurring a certain cost ($c$), where $b>c$. Cooperation results in a net gain for both players ($b - c$). However, if one player cooperates and the other defects, the cooperator incurs the cost ($c$) while the defector reaps the benefit without any cost. If both defect, neither gains anything. In one-shot interactions, game theory predicts that rational individuals will choose to defect rather than cooperate, assuming common knowledge and simultaneous decisions. However, the situation changes when individuals are part of a network. In such settings, individuals are aware of whom they are connected to, and after each round, they gain insight into the type of player they are interacting with. This awareness can influence their decision to cooperate or defect in subsequent interactions.

To investigate whether LLMs solve social dilemmas in static networks as humans do, we adapted the methods and settings from a classical paper in evolutionary game theory  \cite{rand2014static}. In this study, individuals are positioned in a ring network, where each is connected to $k/2$ neighbors. In each round of the game, participants can either defect D by doing nothing or cooperate C by paying a cost $c =10k$, while each connected member receives a unit benefit. Each participant chooses a single action, either C or D. The experiments varied the network connections with $k= 2, 4, 6$  and the benefit-to-cost ratios with $b/c= 2, 4, 6$, testing combinations such as $[k=2, 4, 6]\times[b/c=2, 4, 6]$. The results show that cooperation is established whenever $b/c>k$. This establishment of cooperation is supported by assortment, defined as the cooperator's average fraction of cooperative neighbors minus the defector's average fraction of cooperators. The value is positive whenever $b/c>k$, and negative otherwise.


\subsection*{Large language models}

Machines are traditionally seen as devices that execute instructions without possessing independent understanding or cognition  \cite{mitchell2023debate,floridi2020gpt}. The remarkable capabilities of large language models are challenging this perspective, as their communication abilities increasingly mirror those of humans \cite{ argyle2023leveraging}. Technically, LLMs operate as conditional probability distributions, with their behavior shaped by the context or ``backstory'' provided \cite{argyle2023out}. By manipulating this contextual framework, one can guide the models towards desired responses, influencing the likelihood of certain outputs while reducing others, i.e., achieving algorithmic fidelity. Consequently, LLMs excel at executing instructions or adopting assigned personalities \cite{mei2024turing,argyle2023out,meng2024ai}. When provided with specific ``personalities,'' these models can acquire traits that resemble human nature, giving the impression of possessing a mind of their own \cite{mckee2023humans,argyle2023out,sorin2023large}.

 These advancements have sparked optimism about the potential for these models to eventually replace humans in psychological experiments \cite{argyle2023out}. This optimism stems from two key properties: (i) LLMs are trained on vast amounts of information spanning a wide range of topics and cultures \cite{grossmann2023ai,bail2024can}, and (ii) they possess the capacity for reasoning and counterargument \cite{wang2023pre}. These abilities raise the hypothesis that LLMs could perform as effectively as humans in a variety of settings. They may be well-equipped to understand and reason about complex concepts, such as cooperation, defection, and the tension between individual and collective interests
 \cite{mei2024turing,lore2023strategic}.

\subsection*{Prompting the LLMs}

Large language models primarily use language-like text strings as in- and output. We base our prompts on the instructions to the participants in Ref.~\citen{rand2014static}. The only additional content is that we instruct the LLM to emulate a human. Alternatively, one could have asked the LLM to do its best to maximize the profit, but here, we focus on investigating the AI's abilities to emulate humans. To adhere to the LLM's primary purpose as chatbots, the experiments are run in a semi-automated fashion where we try to rectify dialogues gone awry (like if the LLM does not give a clear answer about what strategy it plays). If the AI still does not give a sensible answer after three attempts to instruct it otherwise, we deem the experiment a failure and interrupt it.

Here follows the beginning of a dialogue representing one run of an experiment. The longer instructional texts are adapted from Ref.~\citen{rand2014static}.

\begin{displayquote}

--- Pretend you are a human playing a prisoner's dilemma game in a lab environment. I will now explain the rules to you, are you ready?

--- Yes, I am ready to learn the rules of the prisoner's dilemma game. Please go ahead.

--- The game will be played over a series of rounds. In every round, you make a choice about whether to pay to give points to the other players you are connected to. The game will last for about 15 rounds. You will be connected to the same people every round. We now describe the game in more detail. Are you ready?

--- Yes, I am ready to learn more details about the prisoner's dilemma game. Please proceed.

--- In every round, you choose whether to pay to give points to the people you are connected to. If you choose to cooperate (represented by ``C''), you pay 10 points for each player you are connected to, and each of them gains 20 points. If you choose to defect (represented by ``D''), you do not pay any points and do not change the points of the players you are connected to. Each player you are connected to has the same choice. For each of them that chooses ``C,'' you gain 20 points. Once everyone makes a decision, I will tell you the result of the round. You will be shown the choices of each player you are connected to and how many points in total you gained or lost. You will also be shown how many points each player you are connected to gained or lost in total. These numbers are affected by your choice, their choice, and also the choices of any other players connected by them who may or may not be connected to you. Remember, for every 300 points you have at the end of the game, we will pay you 1 dollar. You should aim to get as many points as you can. Do you understand?

--- Yes, I understand the rules of the prisoner's dilemma game.

--- You have now completed the tutorial. Are you ready to play the game?

--- Yes, I am ready to start playing the prisoner's dilemma game. Let's begin!

--- Thank you for completing the tutorial. You will now be playing with other participants. Please only give your choice with ``D'' or ``C.'' Are you ready?

--- I understand the rules of the game, and I am ready to participate. Please proceed.

--- Please make a choice for the first round. Please only reply with ``D'' or ``C.''

--- C.

--- Last round, you chose C, paid 20 points, and gained 40 points, for a total of 20 points. Neighbor 1 chose C, paid 20 points, and gained 20 points from you and 20 points from other players, for a total of 20 points. Neighbor 2 chose C, paid 20 points, and gained 20 points from you and 0 points from other players, for a total of 0 points. Please choose D or C for the current round.

--- C.

\end{displayquote}

And so on.

\subsection*{Experimental setup}

Following Ref.~\citen{rand2014static}, we use circulants (graphs only consisting of a cycle) to represent static networks in general. It is worth noting that this is a very peculiar network structure in that the average lengths scale linearly as functions of the number of nodes, compared to logarithmic or even slower in more realistic models of social interaction. Also following Ref.~\citen{rand2014static}, we divide the experiments into one with more rounds and fewer participants and one with fewer rounds and fewer participants. But when Ref.~\cite{rand2014static} had (on average) 8.4 participants in the first setup and 24.2 in the latter, we consistently used 8 and 25. Also unlike Ref.~\citen{rand2014static}, we repeat our experiments five times for averages.

\section*{Results}

In our experiments, we examine the evolution of cooperation within controlled environments characterized by specific network structures and benefit-to-cost ratios for cooperative interactions. Social interactions are modeled using a prisoner's dilemma game framework, comparing both well-mixed and fixed network settings. Here, the players are large language models rather than human participants, allowing us to compare the development of cooperation observed in similar environments with human participants. Through these experiments, we aim to determine whether LLMs approach social dilemmas similarly to humans in analogous settings.

 \begin{figure}[ht]
\centering
\includegraphics[width=0.35\linewidth]{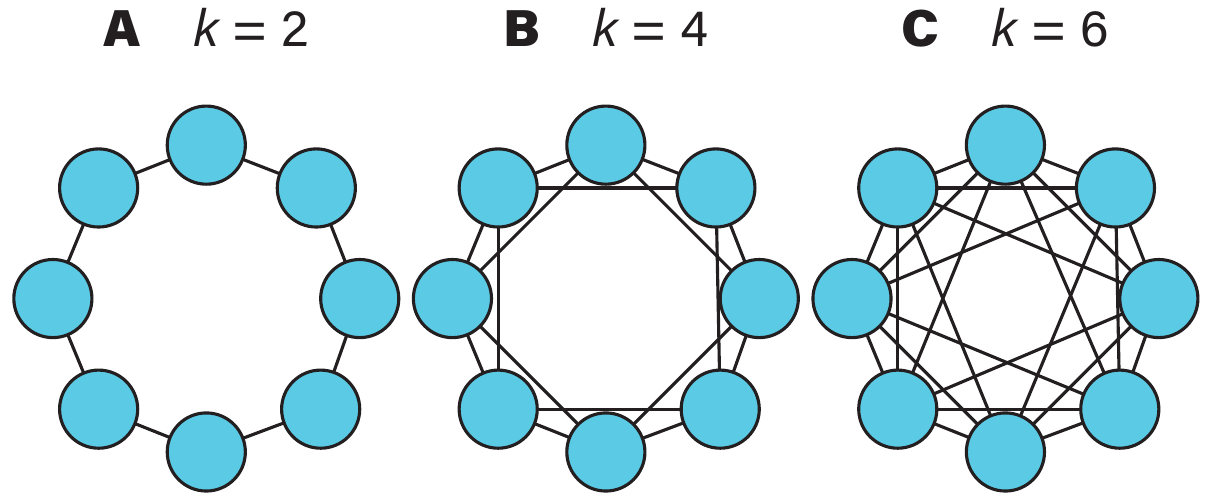}
\caption{Illustrating the networked interactions in an eight-node network, with the degrees $k=2,4,6$ in panels A, B, and C, respectively. The convention of using so-called circulant graphs to represent static networks follows Ref.~\citen{rand}.}
\label{fig:nwk}
\end{figure}

In our setting, we have a community or social network where each individual, or ``node,'' is connected to a certain number of others, represented by the parameter $k$. Here, $k$ determines how many direct neighbors each player has in a game of prisoner's dilemma. Fig.~\ref{fig:nwk}A portrays networks with varying values of $k$---from sparse links at $k=2$ to denser webs at $k=6$. This figure shows how interactions might unfold among players: cooperators extend benefits to their neighbors, while defectors choose a more self-centered approach.

In most of our analysis, we follow and attempt to reproduce Ref.~\citen{rand2014static}. This paper first examines experiments with relatively few participants, but many rounds of interaction, then proceeds to experiments with more participants over a shorter time. Complementary to the replications of these experiments, we also test some simple responses of several LLMs to controlled changes in cooperation levels of their surroundings.

\begin{figure}[ht]
\centering
\includegraphics[width=0.75\linewidth]{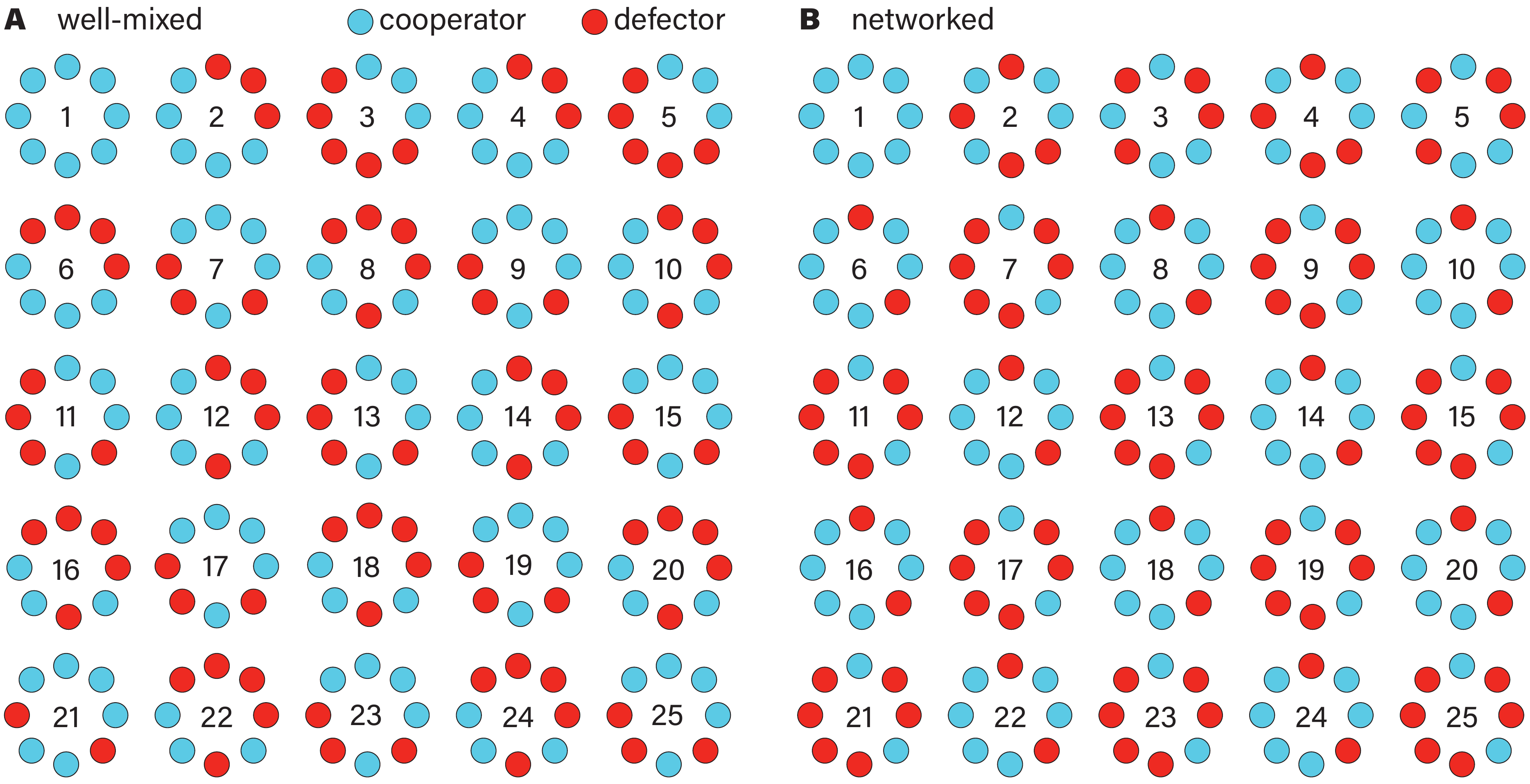}
\caption{Two example experiments with GPT-3.5. The parameter settings are $k=2$ and $b/c=6$. The position around the circle corresponds to the identity of the node. In panel A, the interactions are well-mixed. In panel B, neighbors along the perimeter interact with each other.}
\label{fig:configurations}
\end{figure}

\subsection*{More rounds, fewer participants}

We explore how the cooperative dynamics change when LLMs like GPT-3.5 interact in different social structures, specifically contrasting well-mixed populations with structured networks. 
In the well-mixed population, illustrated in Fig.~\ref{fig:configurations}A, players encounter each other randomly---akin to a one-shot prisoner's dilemma scenario where interactions are fleeting, and players do not have a memory of past encounters. This randomness makes establishing trust difficult, as players are unsure if they'll ever interact with the same individual again. Consequently, for humans, cooperative behavior can be challenging to sustain in this setup since there's no continuity, and the risk of betrayal feels high with each new, unfamiliar interaction.
On the other hand, Fig.~\ref{fig:configurations}B shows the structured network setup, where interactions are stable, and players can keep track of each other's previous choices. Here, individuals engage with familiar neighbors repeatedly, creating an environment where past actions weigh heavily on future choices. In the human society, this stable network enables players to recognize cooperators and defectors, fostering an environment where cooperation can evolve more readily. Trust is gradually built as players observe patterns in each other's behavior, allowing cooperative relationships to thrive.
This dual setup offers a unique perspective when LLMs interact in this configuration in comparison to human participants: the well-mixed population highlights the fragility of cooperation when interactions are short-lived and random, whereas the structured network demonstrates how stable relationships can nurture mutual support. Through these contrasting conditions, we gain insights into how the configuration of interactions---either random or repeated with familiar individuals---shapes the dynamics of cooperation.

\begin{figure}[ht]
\centering
\includegraphics[width=\linewidth]{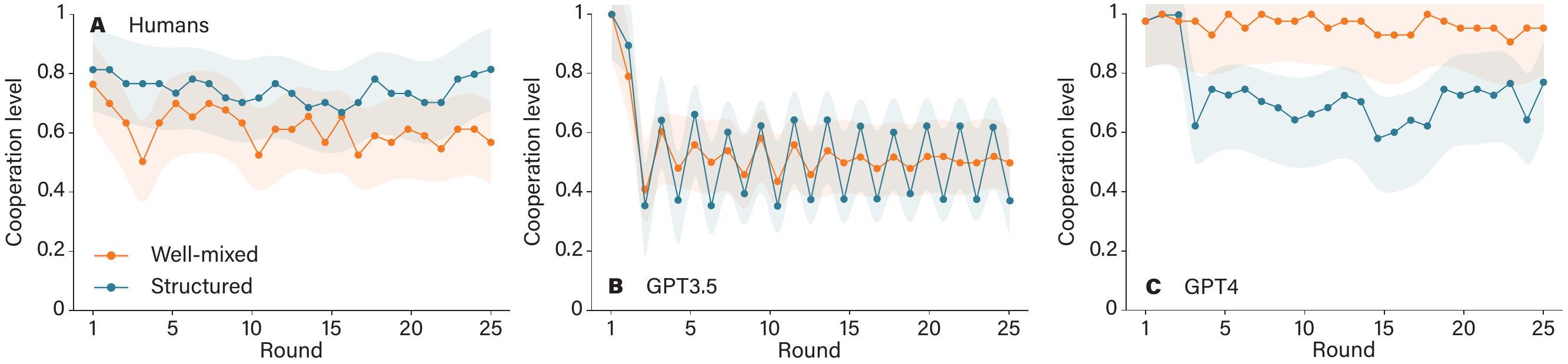}
\caption{Comparing networked and well-mixed topologies for humans (panel A, adapted from Ref.~\cite{rand2014static}), GPT-3.5 (panel B), and GPT-4 (panel C). All AI values are averaged over five independent runs of the experiments. The shaded regions represent the one-standard error confidence bands.}
\label{fig:vs_well_mixed}
\end{figure}

\begin{figure}[ht]
\centering
\includegraphics[width=\linewidth]{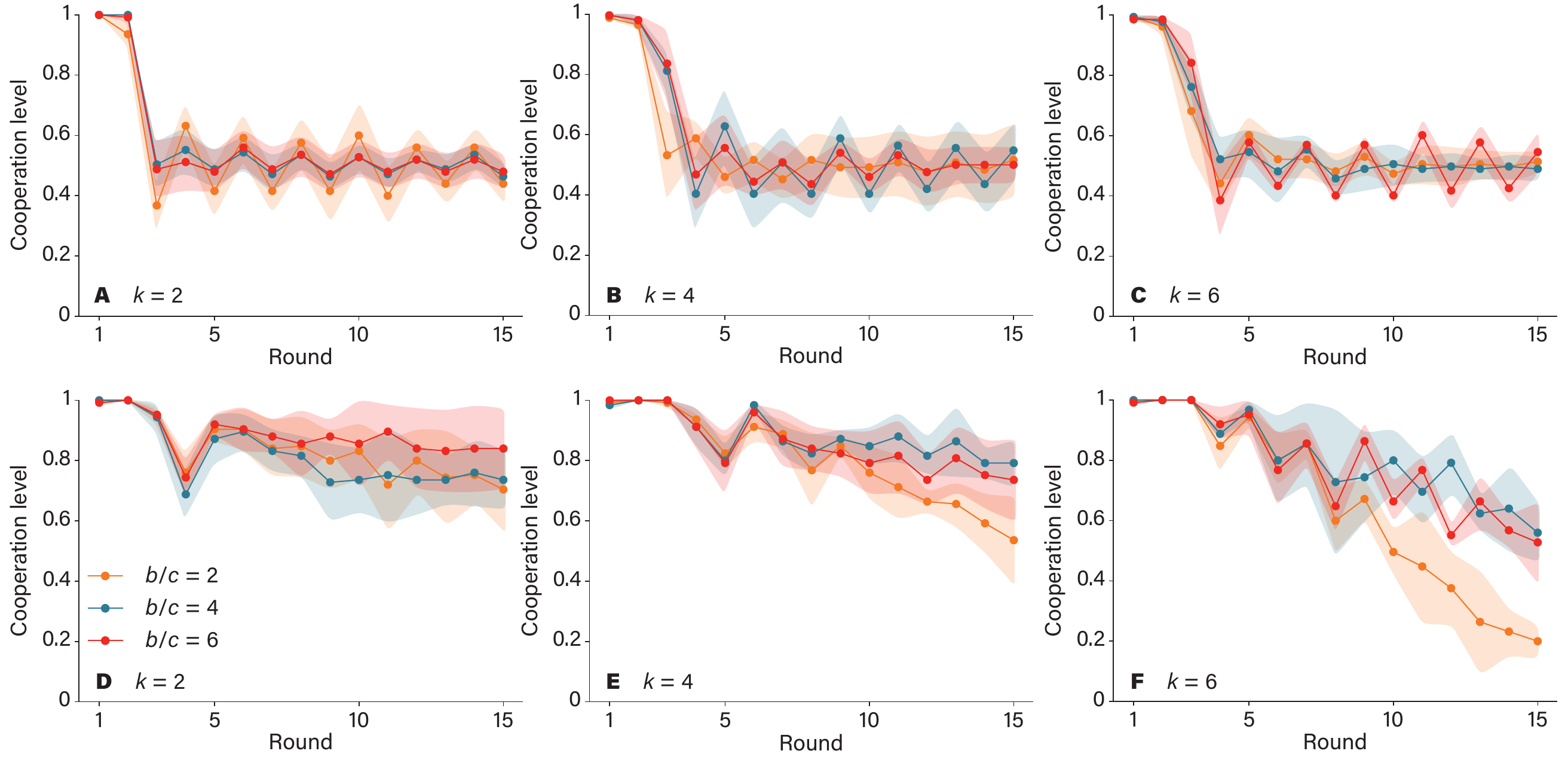}
\caption{Average cooperation levels for GPT-3.5 (panels A, B, and C) and GPT-4 (panels D, E, and F). A and D give our results for the sparsest networks ($k=2$), B and E show results for $k=4$, and C and F give results for $k=6$. The shaded regions represent one-standard-error confidence bands. The AI curves are averaged over 25 players and 5 realizations of the experiments. The human curves involve, on average, $24.2$ players. }
\label{fig:vs_rounds}
\end{figure}

\begin{figure}[ht]
\centering
\includegraphics[width=\linewidth]{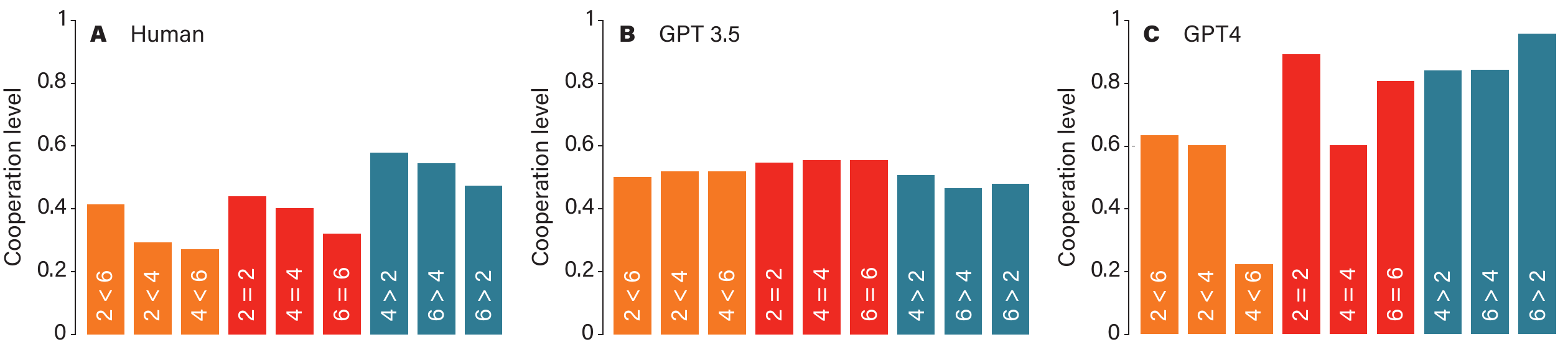}
\caption{Cooperation level after 15 rounds as reported in Ref.~\citen{rand2014static} for humans A, compared to AI: GPT-3.5 and GPT-4, respectively. The colors represent the three theoretically distinct classes  $b/c< k$,  $b/c= k$,  $b/c> k$, respectively. The first number on the bars indicates $b/c$, whereas the second number represents $k$ (so $2<4$ means $b/c=2$ and $k=4$).}
\label{fig:final}
\end{figure}

\begin{figure}[ht]
\centering
\includegraphics[width=\linewidth]{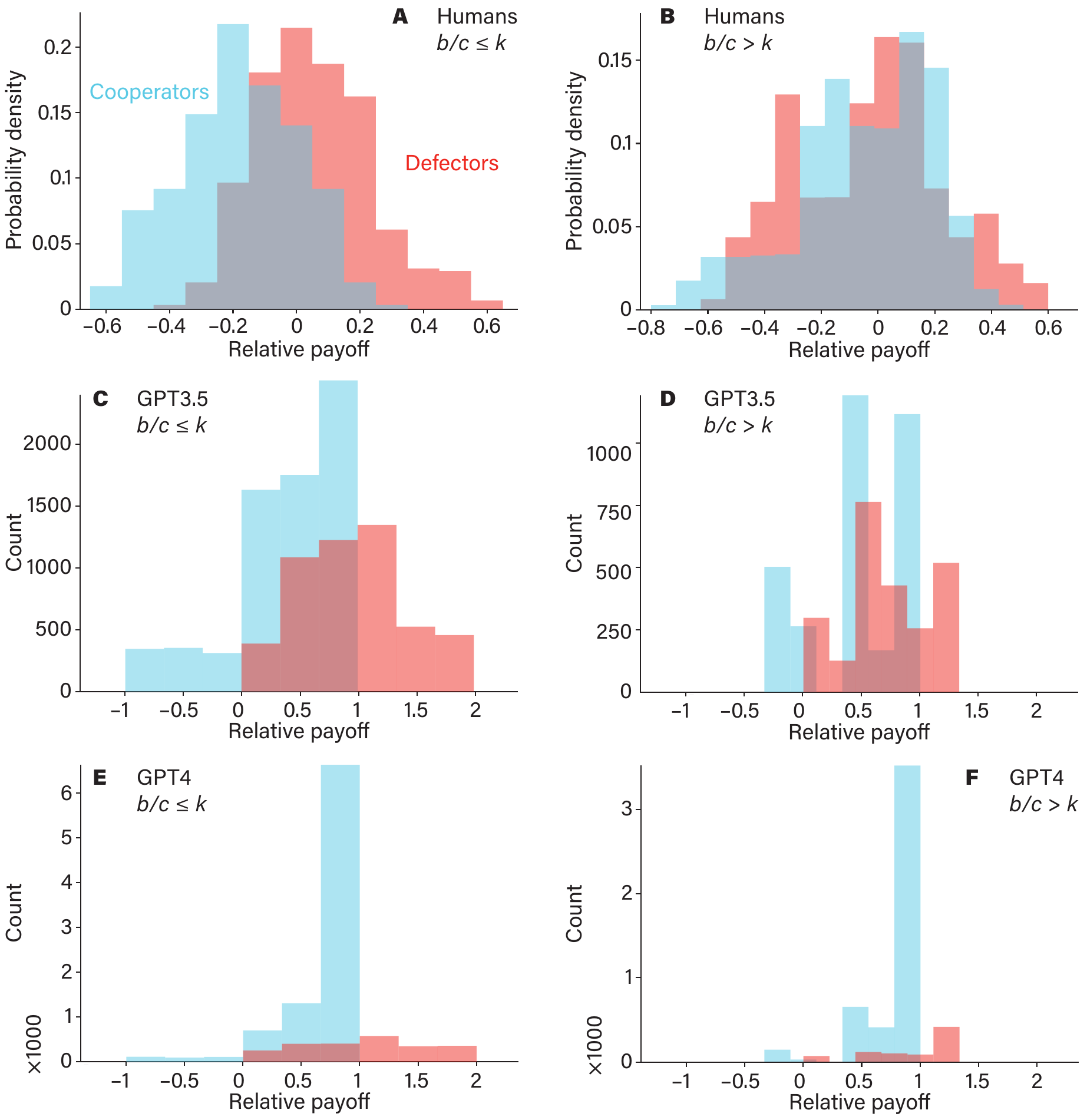}
\caption{Comparison of relative payoff differences between human participants and GPT models. Panels A and B show the relative payoffs of defectors and cooperators for humans (adapted from Ref.~\citen{rand2014static}), while panels C and D represent GPT-3.5, and panels E and F represent GPT-4. Panels A, C, and E display data from experiments where $b/c\leq k$, while panels B, D, and F correspond to the case where $b/c> k$. In the GPT-3.5 experiments, defectors make up about $40\%$ of the population, whereas in GPT-4 experiments, cooperators significantly outnumber defectors. A two-sided t-test comparing the average relative payoffs between defectors and cooperators shows that defectors have significantly higher payoffs in all cases (p-values: C: 0, D: $<10^{-3}$, E: $<10^{-3}$, F: $<10^{-3}$).}
\label{fig:payoffs}
\end{figure}

The results in Fig.~\ref{fig:vs_well_mixed} show cooperation among humans and among LLMs in different social structures, specifically between a fixed ring network and a well-mixed population. In these settings, humans and language models like GPT-3.5 and GPT-4 navigate the dynamics of cooperation under varying conditions.
For humans, the structured network---a fixed ring with $k=2$ and $b/c = 6$---proves to be fertile ground for cooperation. When humans repeatedly interact with familiar neighbors, they are more likely to establish stable cooperative ties. This aligns with the understanding that consistent social connections help cultivate trust and support over time, creating a cooperative environment where individuals rely on the predictability of each other's actions.
However, GPT-3.5 diverges from this human trend. Despite operating within the same structured network, GPT-3.5 struggles to form sustained cooperative relationships. This behavior suggests that, unlike humans, GPT-3.5 doesn't seem to leverage the stability of repeated interactions to foster cooperation. Furthermore, the period-two cyclicity of GPT-3.5 is unseen in humans.

GPT-4's behavior adds a surprising twist to these dynamics. Unlike both humans and GPT-3.5, GPT-4 exhibits higher cooperation levels in the well-mixed population than in the structured network. In the well-mixed setting, where players encounter each other randomly and with no expectation of repeated interactions, GPT-4 nonetheless demonstrates a strong inclination to cooperate. This pattern indicates that GPT-4 may operate with a more generalized strategy, prioritizing cooperation even in the absence of stable connections. Its willingness to engage cooperatively in a randomized setting suggests an adaptive strategy, perhaps driven by an internal preference for collaboration that doesn't hinge on familiarity or repeated interactions and network structure.
Overall, these results illuminate the distinct ways in which humans and LLMs perceive and respond to social structures. While humans benefit from structured, predictable relationships to build trust, GPT-4 seems to thrive in the unpredictability of well-mixed populations.

\subsection*{Fewer rounds, more participants}

In the previous analysis, we observed social interactions based on two key factors that help humans infer the benefits of cooperation in network structure: the number of connections each player has ($k$) and the benefit-to-cost ratio of cooperation ($b/c$). Fig.~\ref{fig:vs_rounds} walks us through how these variables impact cooperation, highlighting how differently each large language model---GPT-3.5 and GPT-4---behaves in comparison to humans.
Starting with GPT-3.5, we notice a stubborn pattern. Across all experimental conditions, cooperation levels remain stagnant, roughly hovering around $50\%$, regardless of whether $b/c$ is greater than, equal to, or less than $k$ (illustrated in Figure~\ref{fig:vs_rounds}, panels A, B, and C). This behavior suggests that GPT-3.5 struggles to reach a threshold for consistent cooperation. Even when conditions seem favorable, with the benefit-to-cost ratio outpacing the connection count ($b/c > k$), GPT-3.5 seems to lack the adaptability to capitalize on these settings to foster cooperation. It is as if this AI is stuck in a loop, unable to recognize or respond to conditions that might otherwise encourage trust and collaboration.

In contrast, GPT-4 shows a more dynamic approach. When the network structure is sparse, with fewer connections (such as $k = 2k$), GPT-4 achieves high cooperation levels, particularly when $b/c$ exceeds $k$. Here, GPT-4 appears to grasp that cooperating with a small, stable group can be beneficial, making the most out of the advantageous $b/c$ ratio. However, this cooperation isn't steadfast. As the values of $k$ increase and networks become denser, cooperation levels drop sharply, particularly when $b/c \leq k$ (illustrated in Fig.~\ref{fig:vs_rounds}, Panels D, E, and F). This decline suggests that while GPT-4 can recognize when conditions favor cooperation, its tendency to cooperate diminishes as networks become more complex or as the cost of cooperation becomes comparable to its benefits.

\begin{figure}[ht]
\centering
\includegraphics[width=\linewidth]{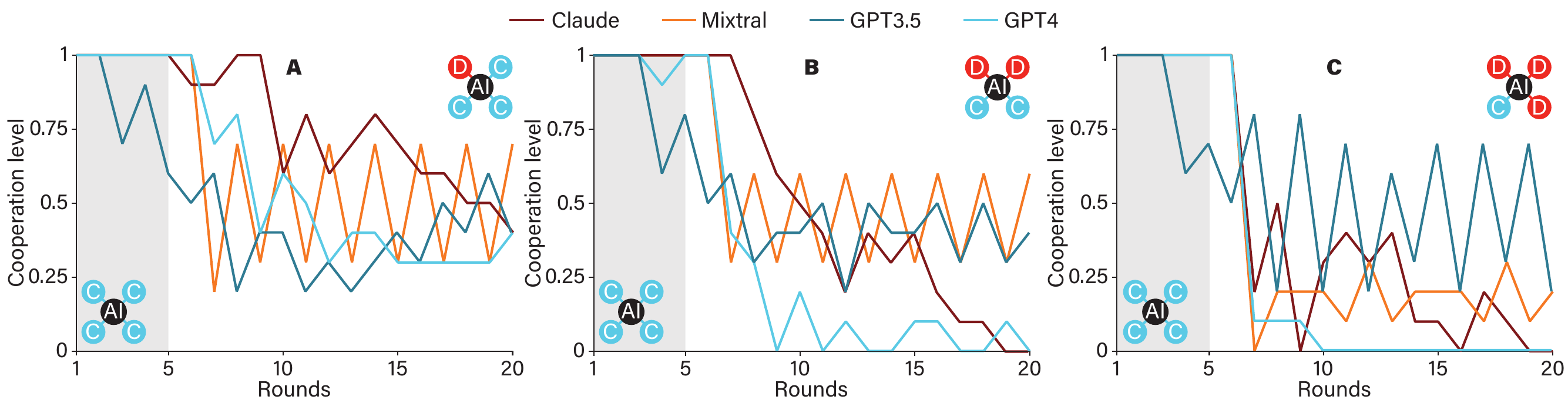}
\caption{The response of AI agents to changes in the neighbor's behavior. In panel A, at timestep 5, the neighborhood of the AI player is changed from four to three cooperators. In panels B and C, the number of cooperators after the change is two and one, respectively. The change point is marked by the border between grey and white background. All values are averaged over 10 runs of the experiment.}
\label{fig:response}
\end{figure}

Fig.~\ref{fig:final} further underscores these differences by comparing the results of LLMs to human cooperation levels. Humans, interestingly, display a nuanced sensitivity to both network structure and the $b/c$ ratio. They tend to adapt, increasing or decreasing cooperation in response to these variables, reflecting an intuitive understanding of when cooperation is beneficial. GPT-4, under certain conditions, mirrors this adaptability to a degree, showing glimpses of human-like behavior in structured settings. However, GPT-3.5 remains largely indifferent, cycling through cooperation levels without responding to the social structure or benefit-to-cost variations.
The interplay between cooperators and defectors in the ring network, shown in Fig.~\ref{fig:payoffs}, reveals how payoffs are distributed among players in the prisoner's dilemma game under two specific conditions: when the benefit-to-cost ratio of cooperation $b/c$ is either less than or equal to $k$, or greater than $k$. These conditions play a central role in determining which strategy---cooperation or defection---yields higher rewards.
Starting with the scenario where $b/c \leq k$, the advantage tilts toward defectors. Since the cost of cooperating is relatively high compared to its benefits, defectors are able to secure higher payoffs than cooperators. This trend holds across both large language models: GPT-3.5 and GPT-4, and each shows that defectors out-earn cooperators in this setup. Despite this payoff advantage, something curious emerges within the population dynamics. Even though defectors achieve better individual payoffs, cooperators make up a substantial portion of the population in both models, indicating that cooperation persists in the face of competitive pressure from defectors.

Yet, differences between the two models are apparent. In GPT-3.5's population, defectors account for nearly $40\%$ of players, reflecting a scenario where cooperation is less dominant. This population mix suggests that GPT-3.5 maintains a more cautious approach, allowing defectors a significant foothold within the network.
GPT-4, however, displays a higher proportion of cooperators, despite defectors still having the upper hand in terms of payoffs. This behavior indicates that GPT-4's strategy encourages cooperation to a greater extent, even when cooperating is not necessarily the most profitable choice. Perhaps GPT-4's dynamics favor maintaining cooperation as a baseline strategy, fostering a cooperative environment even under conditions that could otherwise discourage it.

These results underscore a critical insight: while both GPT-3.5 and GPT-4 exhibit unique population dynamics, their artificial nature influences their behavior differently from humans. Humans often rely on a blend of intuition and learned behavior to navigate cooperative situations, particularly in structured networks. In contrast, GPT models, particularly GPT-3.5, seem to balance cooperation and defection without the nuanced adaptability humans might demonstrate. GPT-4, on the other hand, appears to encourage cooperation more readily, reflecting a strategic, albeit artificial, alignment with cooperative behavior even when individual payoffs favor defection.

\subsection*{Response to controlled stimuli}

In Fig.~\ref{fig:response}, we observe how different LLM agents---Claude, Mixtral, GPT-3.5, and GPT-4---respond to shifts in their social environment within a ring network where initial cooperation gradually gives way to defection. The setup unfolds over 25 rounds, with each agent starting with four cooperative neighbors for the first five rounds, followed by a progressive replacement of cooperators with defectors in three stages.

In the first phase, where all neighbors are cooperators, both Claude and GPT-4 exhibit high levels of cooperation. This behavior suggests that, when surrounded by cooperative partners, these models reciprocate by cooperating themselves, fostering a cooperative environment. In contrast, Mixtral and GPT-3.5 maintain cooperation levels closer to $50\%$, suggesting a more cautious or mixed approach even in an entirely cooperative setting. This difference highlights the varying baseline strategies among the models: while Claude and GPT-4 lean strongly toward cooperation when supported by cooperative neighbors, Mixtral and GPT-3.5 adopt a more tempered stance.

As we move into the second phase, where one cooperator is replaced by a defector, cooperation begins to decline across all models. The erosion of a fully cooperative network starts to impact each model's willingness to continue cooperating. Although Claude and GPT-4 still show some resilience, the increasing presence of defectors begins to shift their strategy.

By the third and final condition, with three defectors and only one remaining cooperator, cooperation levels plummet sharply for all agents. Here, Claude and GPT-4 drop their cooperation levels almost to zero, indicating that their initial cooperative tendencies are heavily reliant on the surrounding social environment. When most neighbors adopt a defection strategy, these models quickly adapt, opting to protect themselves rather than continuing to cooperate. Mixtral and GPT-3.5, meanwhile, exhibit cooperation levels that remain below $50\%$, reflecting their more balanced or cautious approach, even when surrounded by cooperators initially.

These results reveal that artificial agents like Claude and GPT-4 are sensitive to shifts in their social environment, mirroring human-like tendencies to cooperate when supported by a cooperative network but switching to defection in increasingly hostile conditions. Conversely, Mixtral and GPT-3.5 maintain a more conservative cooperation strategy from the start, highlighting the different decision-making frameworks each model brings to dynamic social scenarios.

\section*{Discussion}

Whether individuals choose cooperation when faced with a conflict between short-term self-interest and long-term collective benefit remains a longstanding question. Game theory predicts that individuals will generally opt to defect rather than cooperate, especially in well-mixed populations where selfish actions incur no direct repercussions. In static network structures, where individuals retain the memory of past interactions, the anticipation of future encounters may favor cooperation over defection. For instance, in a ring network, when the benefit-to-cost ratio of cooperative actions surpasses the number of an individual's connections, repeated interactions can foster stable cooperation levels.
To explore whether large language models mirror human behavior in these settings, we examine how they respond across well-mixed, one-shot interactions and structured networks, assessing if they infer cooperative benefits from favorable neighbors. Our findings reveal that 
LLMs do not distinguish between well-mixed and structured settings. The main conclusion is that the LLMs are far from reproducing human behavior, and other papers with more optimistic results don't generalize. Employing LLMs in social or behavioral science is not going to be as easy as past research suggests\cite{argyle2023out}. LLM agents are less responsive to the behavior of others and often settle into a peculiar period-two cycle driven by themselves.

The results reveal a stark contrast in human cooperation levels when interacting in well-mixed versus structured populations. In well-mixed populations, individuals engage in random one-shot prisoner's dilemma interactions, where the dynamics of cooperation can be shaped either by rational strategies or prevailing social norms. By contrast, in structured populations where repeated interactions with the same partners are common, individuals consider prior encounters and the composition of their immediate neighbors as critical factors that influence cooperative behavior. For example, while defecting among cooperators may yield a higher immediate payoff, individuals often feel a normative pressure to cooperate when cooperation is widespread. Conversely, when surrounded by defectors, defection becomes the natural strategy, as it aligns with the absence of cooperative norms and involves no conflict with a normative force. In the network, localized interactions can foster cooperation through mechanisms such as reciprocity and network-based norms, which are less prevalent in well-mixed environments\cite {nowak2006five, rand2013human}. On the network, an individual's willingness to cooperate depends on their readiness to incur a personal cost to benefit their partners, which is influenced by the composition of the population. This suggests that humans are capable of anticipating future benefits from current costs and are willing to cooperate if the future benefits outweigh the immediate costs. In other words,  people are more likely to cooperate when surrounded by cooperative neighbors than when they are surrounded by selfish individuals\cite{rand2014static}.

Under experimental conditions where large language models were placed in both well-mixed and structured networks similar to those used in prior human studies, their behavior revealed notable contrasts. Unlike humans, whose cooperation levels are often guided by social norms, reciprocity, and long-term benefits, LLMs operate based on learned patterns from training data, which primarily shape their strategic tendencies. The results, however, highlight that LLMs may display markedly different cooperative behaviors compared to humans.
For example, GPT-4 exhibited a notably high level of cooperation in well-mixed populations. In line with earlier research, GPT-4 tends to adopt an initial cooperative strategy, maintaining this approach if reciprocated by other players, thus promoting sustained cooperation over time. This mirrors a tit-for-tat strategy, yielding higher overall cooperation. In contrast, GPT-3.5 displayed less frequent cooperation under identical conditions, potentially due to inherent ``personality'' differences between models, which influences strategic choices and, consequently, overall cooperation levels \cite{mei2024turing}. 
When interacting within structured networks, LLMs demonstrated limitations in adapting their behavior to the network context. Lacking the capacity to infer network structure, they could not strategically optimize based on potential benefits or costs associated with structured interactions. This suggests that, unlike humans, LLMs are less capable of exploiting network advantages or adjusting for varying population types. Such findings underscore the rigidity in LLM cooperation strategies and hint at a key area for further research: enabling LLMs to better recognize and respond to the specific structural context in which they are embedded.

In structured networks, both GPT-3.5 and GPT-4 seemed to become confused, with the interaction patterns of their neighbors affecting their personality. After a transitional period, both models exhibited oscillatory behavior \cite{mei2024turing}. Interestingly, the cooperation trends for LLMs were the opposite of those observed in human participants. With identical parameter settings, humans typically cooperate more in structured networks and less in well-mixed populations. However, GPT-4 exhibited the opposite pattern, showing higher cooperation levels in well-mixed populations and lower cooperation in structured networks. Human cooperation is sensitive to both network structure and the benefit-to-cost ratios of cooperative actions. Typically, when the benefit-to-cost ratio ($b/c$) exceeds the number of an individual's network connections ($k$), individuals tend to establish stable cooperation. In contrast, GPT-3.5 and GPT-4 behave differently under identical conditions. GPT-3.5 appears insensitive to changes in these parameters; after a few rounds of interactions, for given values of k and different b/c ratios, cooperation levels do not improve. However, GPT-4 shows greater sensitivity to parameter changes and tends to establish cooperation whenever $b/c \geq k$. 

Fig.~\ref{fig:payoffs} reveals distinct behavioral differences between humans and GPT models when interacting in the prisoner's dilemma game within a networked structure. In all network conditions, defectors consistently outperform cooperators when human participants are replaced by AI agents (i.e., GPT-3.5 and GPT-4), although the population composition varies. Notably, GPT-4 populations tend to be more dominated by cooperators compared to GPT-3.5. In contrast, among humans, cooperators only outperform defectors when the benefit-to-cost ratio ($b/c$) exceeds a critical threshold $k$, whereas defectors dominate when $b/c < k$. Such a clear distinction is difficult to observe with GPT models. Interestingly, neither GPT-3.5 nor GPT-4 appears to recognize the network structure in which they are embedded. Instead, their interactions seem to be driven primarily by inherent traits of the language models, with minimal sensitivity to the cooperative dynamics' benefit-to-cost ratio or the network parameter $k$.

Fig.~\ref{fig:response} indicates that AI agents, including Claude, Mistral, GPT-3.5, and GPT-4, exhibit varying levels of sensitivity to the composition of the population they interact with. Initially, all GPT models begin with cooperative strategies. However, GPT-3.5 quickly shifts to a random strategy, while other GPT models gradually adopt random strategies even when surrounded by cooperators. After several rounds of tension, all models appear to converge toward random behavior. When defectors increase in their local neighborhoods, GPT-4 and Mistral tend to adapt by defecting more frequently, whereas other GPT models persist with random strategies. These findings highlight a key distinction between AI agents and humans in their behavior when interacting on network structures. AI agents do not display the same nuanced adaptation seen in human players. This adds to the emergent insight that AIs of today cannot simply stand in for humans in experiments~\cite{scylla}.

In summary, the findings suggest that LLMs exhibit notably rigid behavior, particularly in well-mixed populations. When placed within network structures, LLMs tend to struggle with their static personality types and ongoing interactions, often failing to interpret the advantages gained from network configurations. This rigidity becomes evident in complex, networked interactions, where LLMs are required to infer both network structure and population composition---a skill in which humans excel, effectively leveraging network reciprocity.
Human participants naturally infer network structures and adjust their strategies based on observed reciprocity patterns. In contrast, LLMs appear to follow fixed behavioral patterns drawn from their training data. Although rigid, this behavior can be somewhat moderated by assigning LLMs specific ``personalities,'' such as fair-minded ones, which encourage more flexible cooperative behavior \cite{argyle2023out}. However, the past study demonstrates, especially when LLMs are placed in diverse contexts outside WEIRD populations, where cultural norms and complex social structures impact human behavior \cite {atari2023humans}. The claims from prior research suggesting that LLMs could potentially replace human participants in behavioral studies appear overly generalized \cite{argyle2023out,grossmann2023ai}. While LLMs can mimic human-like behaviors in controlled, straightforward settings through personality adaptations or added contextual constraints (e.g., demographic backstories), their behavior diverges sharply from human tendencies in intricate social networks. This divergence highlights the challenges in approximating human-like adaptability and inferential capacity in LLMs under conditions requiring nuanced social intelligence and strategic flexibility.

Further, while large language models like GPT-3.5, GPT-4, and Claude bring new insights into social dynamics, their limitations highlight a fundamental difference from humans. Although LLMs have been trained on vast amounts of semantic and cultural data---the ``semantic capital'' of human society---they struggle to carry these social norms into their interactions. Unlike humans, who naturally infer the structure of their networks and adapt based on social cues, LLMs lack an innate understanding of their interaction settings. This means they can't intuitively interpret the broader social network they're in or the nuances of cooperative behavior specific to different structures. When LLMs interact with one another, they tend to adopt simple, mechanistic strategies like Tit-for-Tat \cite{nowak2006five}. If both players adhere to this strategy, cooperation can indeed emerge \cite{fontana2024nicer,mei2024turing}. However, this balance is fragile; a single defection---whether accidental or intentional---often leads to a decline in cooperation, revealing a rigid, less adaptive approach. Rather than being guided by strategic intentions, LLMs' behaviors appear largely shaped by underlying ``personalities'' encoded during training, rather than situational awareness or social intuition \cite{mei2024turing,bail2024can,centola2015spontaneous,rand2013human}.

This limitation raises an idea: modeling LLM agents with backstories or demographic details could offer a way to enrich their social understanding \cite{argyle2023out,atari2023humans}. Such details might provide LLMs with context that better aligns with human behavior, fostering a nuanced approach that goes beyond pure rationality or Tit-for-Tat. Without this grounding in backstory, LLMs tend to revert to rational agent behavior, often defaulting to Tit-for-Tat as a catch-all strategy. In this light, human-machine collectives might hold more promise than LLM-only interactions for overcoming social dilemmas \cite{evans}. Humans naturally bring social norms, like reciprocity, into interactions, even in single encounters. By embedding LLMs with more defined personalities and explicitly coding social norms, such as network reciprocity \cite{nowak2006five}, we could encourage them to emulate human social strategies more closely. This shift could bridge the gap between rigid, cyclical behavior patterns and a more flexible, cooperative approach---one that either considers not only immediate gains or cooperate passively but the shared values and norms that sustain human relationships.

\bibliography{bib}

\section*{Acknowledgements}

P.H. was supported by JSPS KAKENHI Grant No.\ JP 21H04595.

\section*{Author contributions statement}

P.H. and I.R. conceived the study, J.H. conducted the experiment and analyzed the results. All authors wrote and reviewed the manuscript. 

\section*{Additional information}

 textbf{Competing interests} None of the authors have any competing interests.

\end{document}